\documentstyle[12pt,epsf]{article}
\input{psfig}
\begin{document}

\begin{center}
STRANGENESS PHOTOPRODUCTION WITH THE SAPHIR DETECTOR\footnote{This
work
is supported by the Bundesminister f\"{u}r Forschung und Technologie
(BMFT) and by the
Deutsche Forschungsgemeinschaft (DFG).}

\vspace{.5cm}

\small {J.\ BARTH, M.\ BOCKHORST, W.\ BRAUN, R.\ BURGWINKEL, K.H.\ GLANDER,
S.\
GOERS,
J.~HANNAPPEL, N.\ J\"{O}PEN, U.\ KIRCH, F.\ KLEIN, F.J.\ KLEIN,
\underline{\bf D.\ MENZE},
W.\ NEUERBURG, E.\ PAUL, W.J.\ SCHWILLE, M.-Q.\ TRAN, R.\ WEDEMEYER, F.\
WEHNES,
B.~WIEGERS, F.W.\ WIELAND, J.\ WI{\SS}KIRCHEN}

{\it Physikalisches Institut, Universit\"at Bonn, 53115 Bonn, Germany}

\vspace{6pt}

{J.\ ERNST, H.G.\ J\"{U}NGST, H.\ KALINOWSKY, E.\ KLEMPT,
J.\ LINK, H.v.\ PEE, R.\ PL\"{O}TZKE}

{\it ISKP, Universit\"at Bonn, 53115 Bonn, Germany}

\vspace{6pt}

{M.\ SCHUMACHER, F.\ SMEND}

{\it II. Physikalisches Institut, Universit\"at G\"{o}ttingen,
           37073 G\"ottingen, Germany}

\vspace{6pt}

{T.\ MART}

{\it Jurusan Fisika, FMIPA, Universitas Indonesia, Depok 16424, Indonesia}

\vspace{6pt}

{C.\ BENNHOLD\footnote{Supported by DOE grant DE-FG02-95-ER40907}}

{\it Department of Physics, The George Washington University,
     Washington DC, USA}
\end{center}
\begin{abstract}
\noindent {\small  Statistically improved data of total cross sections and
of angular distributions
for differential cross sections and hyperon recoil polarizations
of the reactions $\gamma$p $\rightarrow K^{+}\Lambda$
and $\gamma$p $\rightarrow K^{+}\Sigma^{0}$ have been collected with the
SAPHIR detector at photon energies between threshold and 2.0 GeV.
Here total cross section data up to 1.5 GeV are presented.
The opposite sign of $\Lambda$ and $\Sigma$
polarization and the change of sign between forward and backward direction
could be confirmed by higher statistics.
A steep threshold behaviour of the $K^{+}\Lambda$ total cross section
is observed.}
\end{abstract}

\noindent {\large \bf 1~~Experimental Data}\\
\noindent Using the SAPHIR detector \cite{1} data of $\gamma p \rightarrow
K^{+} \Lambda$ and
$\gamma p \rightarrow K^{+}\Sigma^{0}$ were taken 
and analyzed.
Starting from 3 reconstructed tracks a p$\pi^{-}$ sub sample was
preselected 
by requesting that the invariant mass of
a track pair
with one positive and one negative charge was within the range of the
$\Lambda$ mass.
From the p and $\pi^{-}$ tracks the secondary vertex of the $\Lambda$ decay
has been determined
while the remaining ($K^{+}$) track was used in addition
to identify the primary vertex.
The $K^{+}\Lambda$ channel was separated from the $K^{+}\Sigma$ one by
kinematical
fits. The separation of
background contributions was achieved by cuts in the missing mass,
in the invariant mass distribution of the $p\pi^{-}$ system and in the
probabilities of the
vertex fits.
Total cross sections,
a complete set of angular distributions of the $\Lambda$ and $\Sigma$
polarization  and differential cross sections for both
reactions have been determined in a photon energy range between threshold
and 2.0 GeV .\\  

\begin{figure}[htb]
\centerline{
\psfig{figure=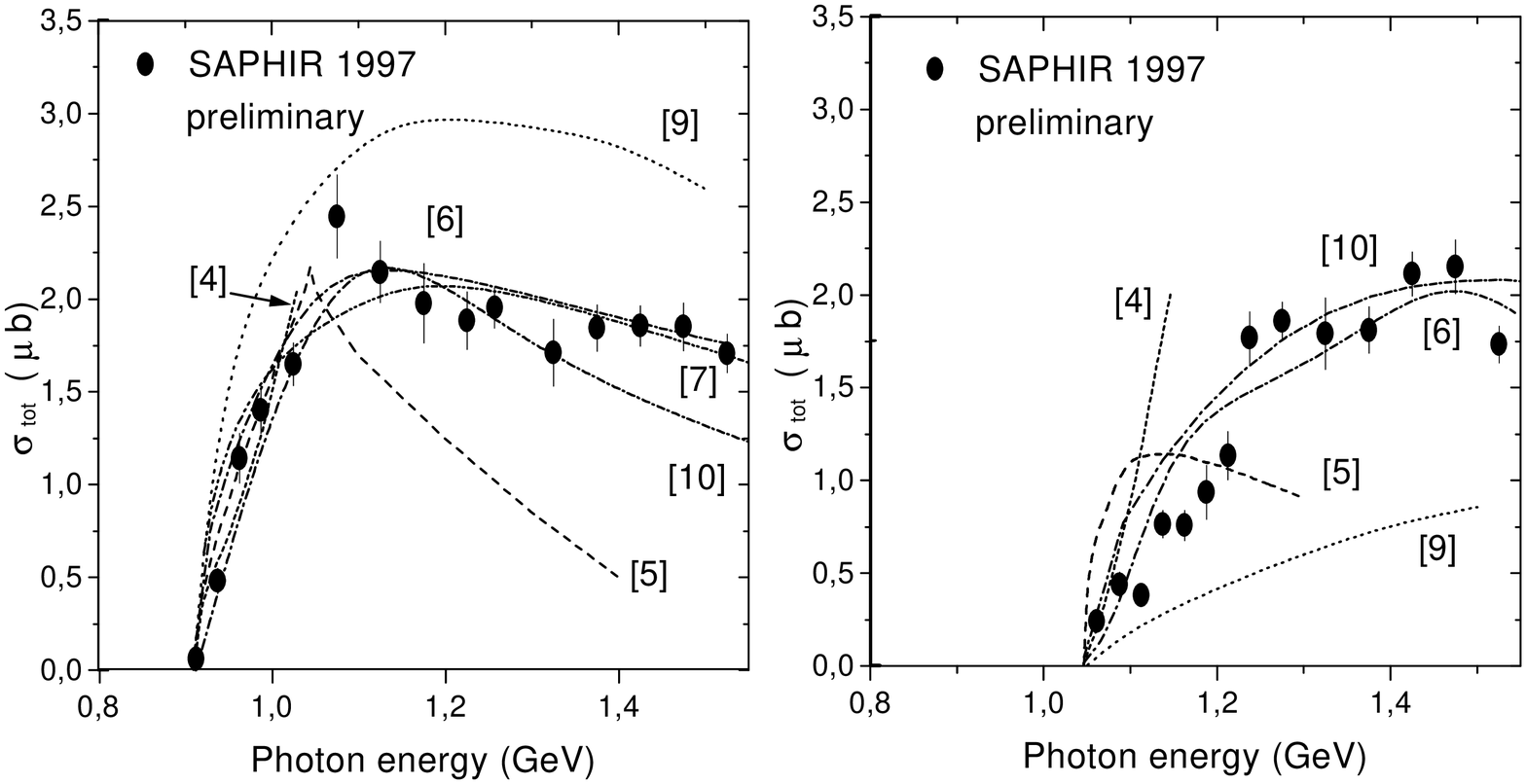,height=7.0cm,width=\textwidth}
}
\caption[]{\footnotesize
\parbox[t]{14.5cm}{Total cross sections for $\gamma p \rightarrow K^{+} 
\Lambda$ (left) and $\gamma p \rightarrow K^{+} \Sigma^{0}$ (right)
in comparison with theo\-re\-ti\-cal calculations. 
Black circles = SAPHIR data \cite{2}. The errors are statistical ones
combined
quadratically with the sytematic errors from the normalization by the
primary photon flux.
The numbers at the curves correspond to the 
references: [4],[5] = chiral calculations, [6],[7] = isobaric models,
[9] = Regge calculations, [10] = quark model.}
}
\label{tot}
\end{figure}

\noindent {\large \bf 2~~Results and Discussions}\\
\noindent The data presented here have been analysed in the course of a
thesis \cite{2}.
They have been taken with a trigger on 2 charged particles in the final
state \cite{1} and 
include the reanalysed data of \cite{3} 
together with additional new data to improve the statistics.
Fig.~\ref{tot} shows the total cross section data in the energy range
between threshold 
and 1.5 GeV.
The analysis of the data at higher energies is still in progress.
The total cross section of the reaction $\gamma p \rightarrow K^{+}
\Lambda$ rises rapidly from threshold up to a pronounced maximum. A
comparison with a chiral 
perturbation theory \cite{4} and with combined channel calculations in
chiral SU(3) 
dynamics \cite{5} shows qualitative agreement in the near threshold region.
In the latter reference the maximum is predicted as a cusp structure near 
the $K^{+}\Sigma^{0}$ threshold \cite{5}. 
While conceptually appealing until now these calculations are limited to
s-wave 
amplitudes.\\
In the case of $K^{+}\Sigma^{0}$ the rise of the total cross section near
threshold is 
smooth and overestimated by both models.

Between threshold and about 1.5 GeV the isobar model constitutes the most
widely used method of
analysis \cite{6,7,8}.
In this phenomenological approach a number of tree-level diagrams
with s-, t- and u-channel resonances
with couplings fitted to the data are included.
The quality of the data until now does not allow to determine
the resonance contributions uniquely; different models include
different sets of resonances.
The problem of overestimating the total cross sections
at energies above 1.5 GeV has been partially solved either by
making use of hadronic form factors \cite{6} or the inclusion
of additional t-channel resonances \cite{7}.\\
A Regge-based model \cite{9}
has been developed that describes
high energy ($E_{\gamma}$ = 6 - 12 GeV)
photoproduction of $\pi$N and KY data. While this description leads to
good agreement with data at very high energies, it severely overestimates
the $K \Lambda$ and underestimates the $K \Sigma$ total cross sections.\\
In addition to the above mentioned models, different quark models
\cite{10,11}
describe the kaon photoproduction cross section data with few parameters.

\begin{figure}[h]
\centerline{
\psfig{figure=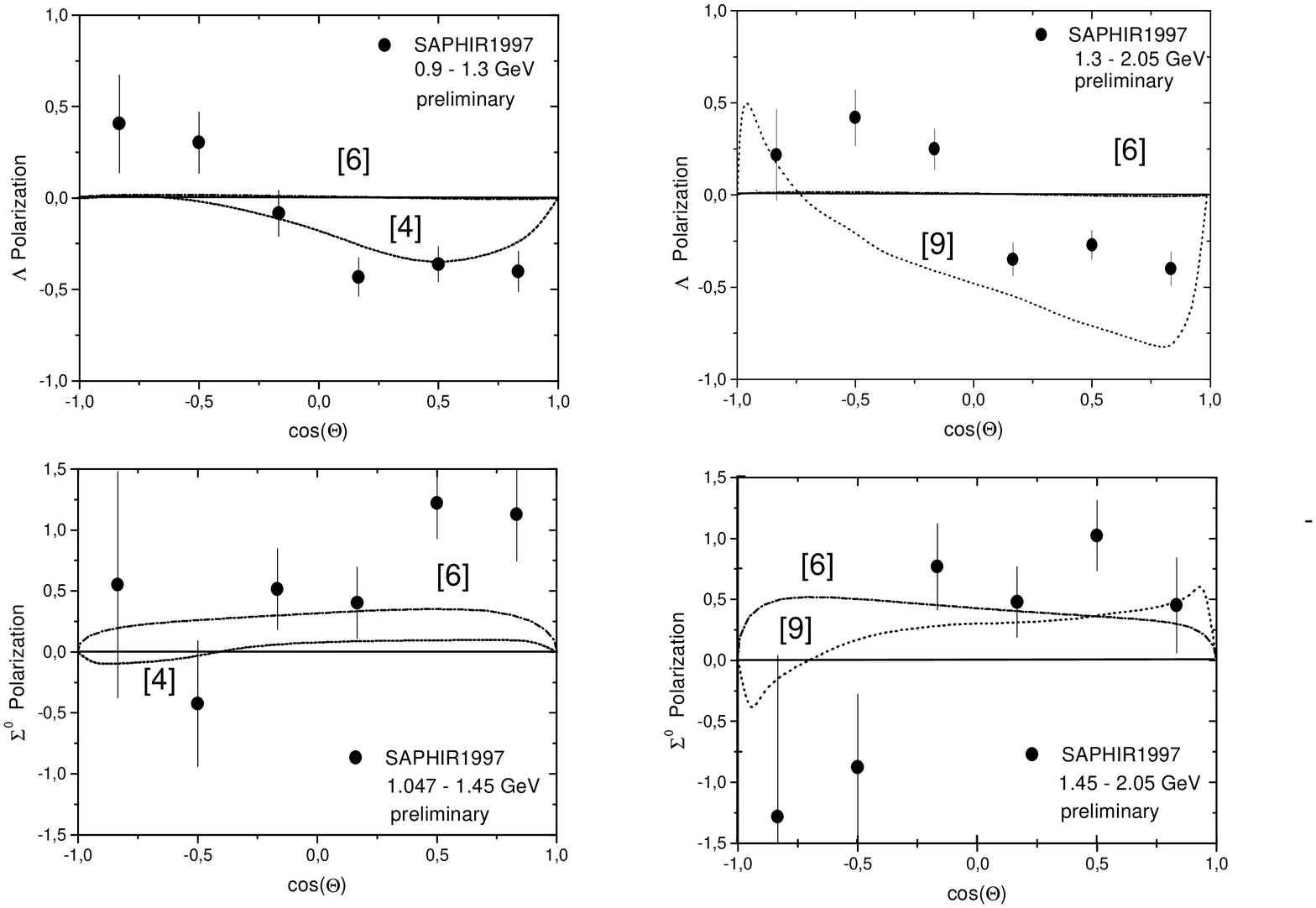,height=7.0cm,width=\textwidth}
}
\caption[]{\footnotesize
\parbox[t]{14.5cm}{Angular distributions of the $\Lambda$
(above) and $\Sigma^{0}$ (below) polarizations,
in dependence on $E_{\gamma}$ (left: lower energies, right: higher
energies)
and in comparison with theoretical calculations.
Black circles = SAPHIR data \cite{2}. The numbers at the curves are
explained in fig. 1.
The prediction of \cite{6} of the $\Lambda$ polarization is so small that
it cannot be distinguished from 
zero in this plot.}
}\label{poll}
\end{figure}

Fig.~\ref{poll} shows the hyperon polarisations in two different energy
bins.
The data show opposite signs of the $\Lambda$ and $\Sigma^{0}$
polarizations. 
Furthermore, there is a sign change between forward and backward direction.

Up to now there is no theoretical model which is able to describe these
features in detail.
The statistics of our data in the near threshold region is not sufficient
to compare them
with chiral calculations \cite{4,5}. In fig.~\ref{poll} the predictions of
\cite{4}
have been extrapolated to 1.25 GeV. Although this is not the threshold
region the 
prediction for the $\Lambda$ polarization shows a negative sign in forward
direction,
for the $\Sigma$ polarization a change in sign between forward and backward
direction.
The extrapolation of the Regge calculations down to 2 GeV also show the
gross features 
of the angular distributions \cite{9}.

\noindent \footnotesize I would like to thank M. Guidal, J.M. Laget, Z. Li
and B. Saghai
for helpful discussions.


\begin{thebibliography}{99}
\footnotesize
\bibitem{1} 
W.J. Schwille et al. , Nucl. Instr. Meth. A {\bf 344}(1994)470.
\bibitem{2}
M.Q. Tran, Ph.D. Thesis, Bonn preprint, BONN-IR-97-11(1997).
\bibitem{3}
SAPHIR collaboration, M. Bockhorst {\it et al.}, Z. Phys. C {\bf 63} 
(1994)37, 
L. Lindemann, Ph.D. Thesis, Bonn preprint, BONN-IR-93-26 (1993),
H. J\"ungst, Ph.D. Thesis, Bonn preprint, BONN-IR-95-19 (1995).
\bibitem{4}
S. Steininger et al., Physics Letters {\bf B391}(1997)446.
\bibitem{5}
N. Kaiser et al., Nucl. Phys. {\bf A612}(1997)297.
\bibitem{6}
T. Mart and C. Bennhold, Nucl. Phys. {\bf A585}(1995)369c,
T. Mart, C. Bennhold, and C.E. Hyde-Wright, Phys. Rev. C{\bf
51}(1995)R1075, 
T. Mart and C. Bennhold, Few-Body Systems Suppl. {\bf 99}(1995)1,
T. Mart, Ph.D. Thesis, Mainz preprint, KPH12/96(1996),
C. Bennhold et al., Washington preprint, Nucl-th/9703004(1997).
\bibitem{7}
R.A. Adelseck and B. Saghai, Phys. Rev. {\bf C42}(1990)108, 
J.C. David {\it et al}., Phys. Rev. {\bf C53}(1996)2613.
\bibitem{8}
R.A. Williams, C.-R. Ji, S.R. Cotanch, Phys. Rev. {\bf C46}(1992)1617.
\bibitem{9}
M. Guidal et al., preprint Saclay, DAPNIA 97-26 (1997).
\bibitem{10}
Z. Li, Phys. Rev. C {\bf 52}, 1648 (1995).
\bibitem{11}
V. Keiner, Z. Phys. {\bf A352}(1995)215.
\end{thebibliography}
\end{document}